\documentclass[12pt,preprint]{aastex}

\usepackage{xspace,graphicx}
\newcommand{\etal}{{et al.}}
\newcommand{\NH}{\mbox{$N_{\rm H}$}}        % Defines NH
\newcommand{\Os}{O$^{+6}$\xspace}
\newcommand{\Oe}{O$^{+7}$\xspace}
\newcommand{\chandra}{{\sl Chandra}\xspace}

\begin{document}
\title{Chandra Observations of MBM12 and Models of the Local Bubble}
\author{R.~K.~Smith}
\affil{NASA/Goddard Space Flight Center}
\affil{Code 662, Greenbelt, MD 20771}
\affil{Department of Physics and Astronomy}
\affil{The Johns Hopkins University}
\affil{3400 North Charles Street, Baltimore, MD 21218}
\email{rsmith@milkyway.gsfc.nasa.gov}  
\author{R.~J.~Edgar, P.~P.~Plucinsky, B.~J.~Wargelin}
\affil{Harvard-Smithsonian Center for Astrophysics}
\affil{60 Garden Street, Cambridge, MA  02138}
\author{P.~E.~Freeman}
\affil{Department of Statistics, Carnegie Mellon University}
\affil{5000 Forbes Ave., Pittsburgh, PA 15213}
\author{B.~A.~Biller}
\affil{Steward Observatory, University of Arizona}
\affil{933 N. Cherry Ave.}
\affil{Tucson, AZ 85721}

\begin{abstract}
\chandra observations toward the nearby molecular cloud MBM12 show
unexpectedly strong and nearly equal foreground \ion{O}{8} and
\ion{O}{7} emission.  As the observed portion of MBM12 is optically
thick at these energies, the emission lines must be formed nearby,
coming either from the Local Bubble (LB) or charge exchange with ions
from the Sun.  Equilibrium models for the LB predict stronger
\ion{O}{7} than \ion{O}{8}, so these results suggest that the LB is
far from equilibrium or a substantial portion of \ion{O}{8} is from
another source, such as charge exchange within the Solar system.
Despite the likely contamination, we can combine our results with
other EUV and X-ray observations to reject LB models which posit a
cool recombining plasma as the source of LB X-rays.
\end{abstract}
\keywords{ISM: bubbles, supernova remnants---plasmas---X-rays: ISM}

\section{Introduction}
After decades of observation, the nature of the diffuse soft ($\sim
1/4$\ keV) X-ray background is still mysterious.  Early observations
of soft X-ray emission towards many sightlines
\citep{BFM68,Bunner69,Davidsen72} showed there is substantial diffuse
emission \citep[see also reviews by][]{TB72,MS90}.  Broad-band
spectral data from proportional counter observations fit a
three-component model: an unabsorbed $10^6$\,K thermal component, an
absorbed $2\times10^6$\,K thermal component, and an absorbed
powerlaw. The latter two components contribute mostly at higher
energies, while most of the 1/4 keV band comes from the local
(i.e. unabsorbed) $10^6$\,K thermal component.

\citet{Williamson74} excluded on physical grounds all then-proposed sources
for the 1/4 keV band emission other than a hot ($\sim 10^6$\,K)
ionized plasma.  Current theories to explain the origin of this
emission still require an ionized plasma, and include: [1] a local
young supernova explosion in a cavity \citep{CA82, EC84}; [2] a series
of supernovae \citep{IH84, SC01}; or [3] an overionized plasma slowly
recombining after substantial adiabatic cooling \citep{BS94, Frisch95,
B96}.  One additional possibility was first suggested by
\citet{Cox98}, who pointed out that charge exchange from the solar
wind might create at least part of the 1/4 keV band emission.

Independent of the observed 1/4 keV band emission, absorption line
measurements to many nearby stars \citep[\protect{\it e.g.}][]{WVV90,
WVVC91} show that we are surrounded by a irregularly-shaped ``cavity''
with very low density.  The standard model for the LB combines these
two observations into the ``displacement'' model
\citep{Sanders77,Snowden90}.  In this picture, the diffuse X-rays come
from an elongated bubble of hot gas with average radius $\sim
100$\,pc, with the sun near the center.  If the LB is filled with hot
($\sim 10^6$\,K) gas in collisional ionization equilibrium (CIE), the
models of the resulting emission fit the observed 1/4 keV band soft
X-ray spectrum.  However, this phenomenological model explains neither
the origin of the hot gas nor the low density region.

Early LB models that attempted to explain both the hot gas and the low
density region \citep{CA82, EC84} modeled it as a $\sim 10^5$\,year
old supernova remnant.  However, this model predicts a large column
density of \ion{O}{6} that is simply not observed along many
sightlines \citep{SC94, Oegerle04}.  Since every oxygen atom passing
through the blast wave needs to pass through this ionization state,
the model predictions are quite robust and are simply not observed.
In addition, the total thermal energy required by the phenomenological
models is between $0.37-1.1\times10^{51}$\,ergs--nearly the entire
kinetic energy of a supernova.  \citet{SC01} considered models with
multiple supernovae which are allowed by the \ion{O}{6} data and
energy considerations, and which roughly fit the observed broad-band
emission.  The recombining plasma model, described in detail in
\citet{B96}, has also been able to qualitatively fit existing
observations.  However, as discussed below we are now able to reject
the basic recombining plasma model by combining our results with other
EUV and X-ray observations.

\section{X-ray Emission from the LB}

The difficulty in measuring the soft X-ray spectrum has limited
further analysis.  A $10^6$\,K plasma with solar abundances, in
equilibrium or not, is line-dominated in the range 0.1-1.0 keV, but
the first observation able to even partially resolve these lines was
only done recently with the Diffuse X-ray Spectrometer (DXS)
\citep{DXS}.  Most of the lines are from L-shell ions of elements from
the third row of the periodic chart: Si, S, Mg (and their neighbor
Ne), and M-shell ions of Fe around 72 eV.  There should also be some
X-ray emission from the K-shell lines of carbon, nitrogen, and oxygen.

Fitting the spectrum requires a spectral emission code for a
collisional plasma, such as described in \citet{SmithAPEC} or
\citet{KaastraCode}.  These codes model the thousands of lines emitted
by the cosmically abundant elements due to collisional excitation.
However, the lines in the observed spectrum are numerous and the
atomic data for the line emission are incomplete and sometimes
inaccurate.  In any event, no calculated spectrum matches
high-resolution observations of the soft X-ray spectrum, such as those
from DXS, and it is not clear if the problem lies in the physical
models or the atomic data\citep{DXS}.  To avoid these problems,
unambiguous and strong emission lines are needed.  In a $10^6$\,K
plasma, the most abundant line-emitting element is oxygen.  In
equilibrium at $10^6$\,K the dominant stage is helium-like \Os, with
trace amounts of \Oe and O$^{+5}$.  However, the hot gas in the LB
need not be in equilibrium.  If (somehow) a recent ($\lesssim
10^5$\,yr) supernova created the local bubble, the gas would still be
ionizing.  Conversely, if the LB is old ($\gtrsim 10^6$\,yr), it
should be recombining.  In either case, the ratios of the \Os and \Oe
ion populations will not be at their equilibrium values.

The strongest \ion{O}{7} and \ion{O}{8} emission lines are in the soft
X-ray range, between 0.5-0.8 keV. \ion{O}{7} has a triplet of lines
from $n=2\rightarrow1$, the so-called forbidden (F) line at 0.561 keV,
resonance (R) line at 0.574 keV, and the intercombination (I) line
(actually two lines) at 0.569 keV, while \ion{O}{8} has a Ly$\alpha$\
transition at 0.654 keV.  These lines are regularly used as plasma
diagnostics in other situations: \citet{Acton78} discussed the
non-equilibrium ionization effects on \ion{O}{7} lines in solar
flares.  \citet{Vedder86}, using data from the {\it Einstein} Focal
Plane Crystal Spectrometer (FPCS), measured the ionization state in
the Cygnus Loop using these lines.  \citet{Gabriel91} discussed the
general use of \ion{O}{7} and \ion{O}{8} emission lines in hot
plasmas, specifically applied to Einstein FPCS observations of the
Puppis remnant.  The goal of our work is to apply these methods,
already used to model supernova remnants, to the specific case of the
LB.

Earlier observations suggested that oxygen emission lines are strong
in the LB.  \citet{Inoue79} detected the \ion{O}{7} F+I+R emission
lines with a gas scintillation proportional counter.
\citet{Schnopper82} and \citet{Rocchia84}, using solid-state Si(Li)
detectors detected \ion{O}{7} as well as line from other ions.  More
recently, a sounding rocket flight of the X-ray Quantum Calorimeter
(XQC) observed a 1 steradian region of the sky at high Galactic
latitude between 0.1-1 keV with an energy resolution of $5-12$\,eV,
and detected both \ion{O}{7} and \ion{O}{8}.  The \ion{O}{7} flux was
$4.8\pm0.8$\,ph\,cm$^{-2}$s$^{-1}$sr$^{-1}$\ (hereafter line units, or
LU) and the \ion{O}{8} flux was $1.6\pm0.4$ LU\citep{XQC02}.  The
XQC's large field of view means that the source of the photons cannot
be determined, but this does represent a useful upper limit to the LB
emission in this direction.

Measuring the emission coming solely from the LB requires observations
of clouds which shadow the external emission.
\citet{SnowdenMBM12}(SMV93) used ROSAT observations of the cloud MBM12
to measure the 3/4 keV emission in the LB, and found a $2\sigma$\
upper limit of 270 counts s$^{-1}$sr$^{-1}$\ in the ROSAT 3/4 keV
band\footnote{For consistency we present all surface brightnesses in
units of steradians, and note that 1 sr =
$1.18\times10^{7}$\,arcmin$^{2}$}. This band includes both the
\ion{O}{7} triplet and the \ion{O}{8} Ly$\alpha$\ line.  They also fit
a ``standard'' $10^6$\,K CIE LB model assuming a pathlength of $\sim
65$\,pc, and found a good match to the observed 1/4 keV emission with
an emission measure of $0.0024$\,cm$^{-6}$\,pc.  This would generate
only $\sim 47$ counts s$^{-1}$\,sr$^{-1}$\ in the 3/4 keV band, well
within the 2$\sigma$\ limit.  For comparison, this model predicts 0.28
LU of \ion{O}{7} line emission, which is the dominant contribution to
the 3/4 keV band emission.

The ROSAT PSPC had insufficient spectral resolution to separate the
\ion{O}{7} and \ion{O}{8} emission lines from each other, the
background continuum, and possible Fe L line emission.  We therefore
used the \chandra ACIS instrument to redo the SMV93 observations with
higher spectral and spatial resolution.  Our results were
unfortunately affected by a large solar flare during part of the
observation, the CTI degradation experienced by the ACIS detectors
early in the \chandra mission, and the somewhat higher than expected
background.  Despite these issues, we were able to clearly detect
strong \ion{O}{7} and \ion{O}{8} emission lines.

\section{Observations and Data Analysis}

MBM12 is a nearby high-latitude molecular cloud ($l, b =159.2^{\circ},
-34^{\circ}$).  The distance to MBM12 is somewhat controversial.
\citet{HBM86} placed it at $\sim 65$\,pc based on absorption line
studies; this is revised to $60\pm30$\,pc using the new Hipparcos
distances for their stars.  However, \citet{Luhman01}, using infrared
photometry and extinction techniques, found a substantially higher
distance of $275\pm65$\,pc.  \citet{Andersson02} found a similar
distance ($360\pm30$\,pc) for most of the extincting material but also
found evidence for some material at $\sim 80$\,pc.  As part of a much
larger survey of \ion{Na}{1} absorption towards nearby stars,
\citet{Lallement03} found foreground dense gas toward stars 90-150 pc
distant in the direction of MBM12.  They also suggested that the
distance discrepancy could be resolved if the molecular cloud MBM12 is
itself behind a nearby dense \ion{H}{1} cloud.  However, since we use
it as an optically thick shadowing target, so long as MBM12 is nearer
than the non-local Galactic sources of 3/4 keV band X-rays that
contribute to the diffuse background the precise distance is
unimportant.  Indeed, the greater distance is interesting because it
would constrain possible sources of diffuse 3/4 keV band X-rays.

\begin{figure}[ht]
\includegraphics[totalheight=2.3in]{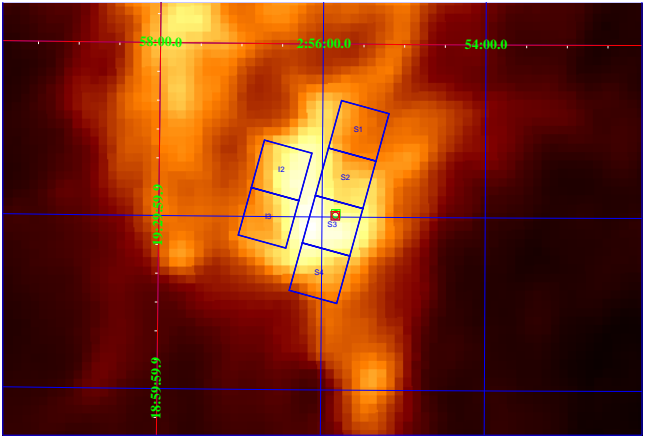}
\caption{IRAS 100\,$\mu$m image of MBM12, with the \chandra observation
coordinates overlaid.  \label{fig:MBM12_IRAS}}
\end{figure}

Our \chandra observations of MBM12 were separated into two nearly
equal sections.  The first observation (hereafter Obi0) was performed
on July 9-10, 2000.  This observation was cut short by a severe solar
flare that led to an automatic shutdown.  Pre-shutdown, the flare
apparently also created a systematically higher background in the ACIS
during the entire observation, which unfortunately meant the entire
observation had to be excluded (see \S\ref{subsec:dp}).  The second
observation (hereafter Obi1), was performed on August 17, 2000 for
$\sim 56$\,ksec during a time of lower solar activity.  The pointing
direction for both observations was $02\fd55\fm50.2,
19\fdg30\farcm14.0\farcs$\ (see Figure~\ref{fig:MBM12_IRAS}).  The
primary CCD, ACIS-S3, was placed on the peak of the 100\,$\mu$m IRAS
emission, which we assume corresponds to the densest part of the
cloud.

%MBM12 is at ecliptic longitude $47.2149\fdg$\ and latitude
%$2.63938\fdg$, almost in the plane of the Solar system.  As discussed
%below, this coincidence may be relevant if some of the emission is due
%to charge exchange onto ions in the solar wind.

MBM12 is a relatively thin molecular cloud and so we must consider if
it is optically thick to background X-rays even at the relatively soft
energies of \ion{O}{7} ($\sim 0.57$\,keV) and \ion{O}{8}
($0.654$\,keV).  Assuming solar abundances, the cross sections at
these energies are $9.4\times10^{-22}$\,cm$^{-2}$\ per H atom and
$6.7\times10^{-22}$\,cm$^{-2}$\ per H atom, respectively \citep{BC92}.
Measuring the column density over the face of the entire cloud is
difficult.  Absorption line measurements show the column density at
one position; \citet{Luhman01} found that most stars in MBM12 have
$A_V < 2$, with background stars in the range $A_V = 3-8$.  Converting
this to an equivalent hydrogen column density via N$_{\rm H}/A_V =
1.9\times10^{21}$\,cm$^{-2}$\,mag$^{-1}$\ \citep{Allen} gives N$_{\rm
H} < 3.8\times10^{21}$\,cm$^{-2}$\ for much of the cloud, with a
maximum value in the range N$_{\rm H} = 6-15\times10^{21}$\,cm$^{-2}$.
SMV93 also measured the column density by cross-correlating the ROSAT
3/4 keV and IRAS 100$\mu$m flux, and derived a value of N$_{\rm
H}/F(100\mu m) = 1.3^{+1.1}_{-0.1}\times10^{20}$\,cm$^{-2}/$(MJy
sr$^{-1}$).  For these observations, the central ACIS-S3 detector was
positioned on the brightest infrared position of MBM12, where
$F(100\mu m) = 30-35$\,MJy sr$^{-1}$, so using the SMV93 ratio,
N$_{\rm H} = 3.6-8.4\times10^{21}$\,cm$^{-2}$.  Since we are observing
in the direction of the densest part of MBM12, we believe N$_{\rm H}
= 4\times10^{21}$\,cm$^{-2}$\ is a conservative value.  The opacity
of MBM12 to background line emission is then $\tau$(\ion{O}{7}) = 3.76 and
$\tau$(\ion{O}{8})= 2.68, and any distant emission will be reduced by
98\% at \ion{O}{7} and 93\% at \ion{O}{8}.  Only an extremely bright
background source could affect our results, and this is unlikely.  The
ROSAT PSPC had a large $2^{\circ}$\ field of view, and although SMV93
found some 3/4 keV emission off-cloud, it was only $\sim
3.3\times$\,brighter than their $2\sigma$\ upper limit for the 3/4 keV
towards the cloud.

\subsection{Data processing\label{subsec:dp}}

The data reduction process was somewhat unusual, since emission from
the LB completely fills the field of view, and the features of
interest are extremely weak.  Our original plan was to use the
front-illuminated (FI) CCDs to measure the individual lines, and the
back-illuminated (BI) CCDs (which have more effective area but lower
resolution and higher background) to confirm the result.  Based on the
\citet{Gendreau} measurement of the soft X-ray background, we expected
pre-launch to obtain at most $\sim 0.004$\ counts/s from \ion{O}{7}
and $\sim 0.001$\ counts/s from \ion{O}{8} lines from both the 4 (FI)
ACIS-I CCDs and the (BI) ACIS S3 CCD.  However, after the proton
damage to the FI CCDs early in the mission, the highly row-dependent
response of the FI CCDs means that these lines would be very difficult
to extract robustly.  Therefore, we focused on data from the BI CCD ACIS-S3.

\begin{figure}[ht]
\includegraphics[totalheight=2.3in]{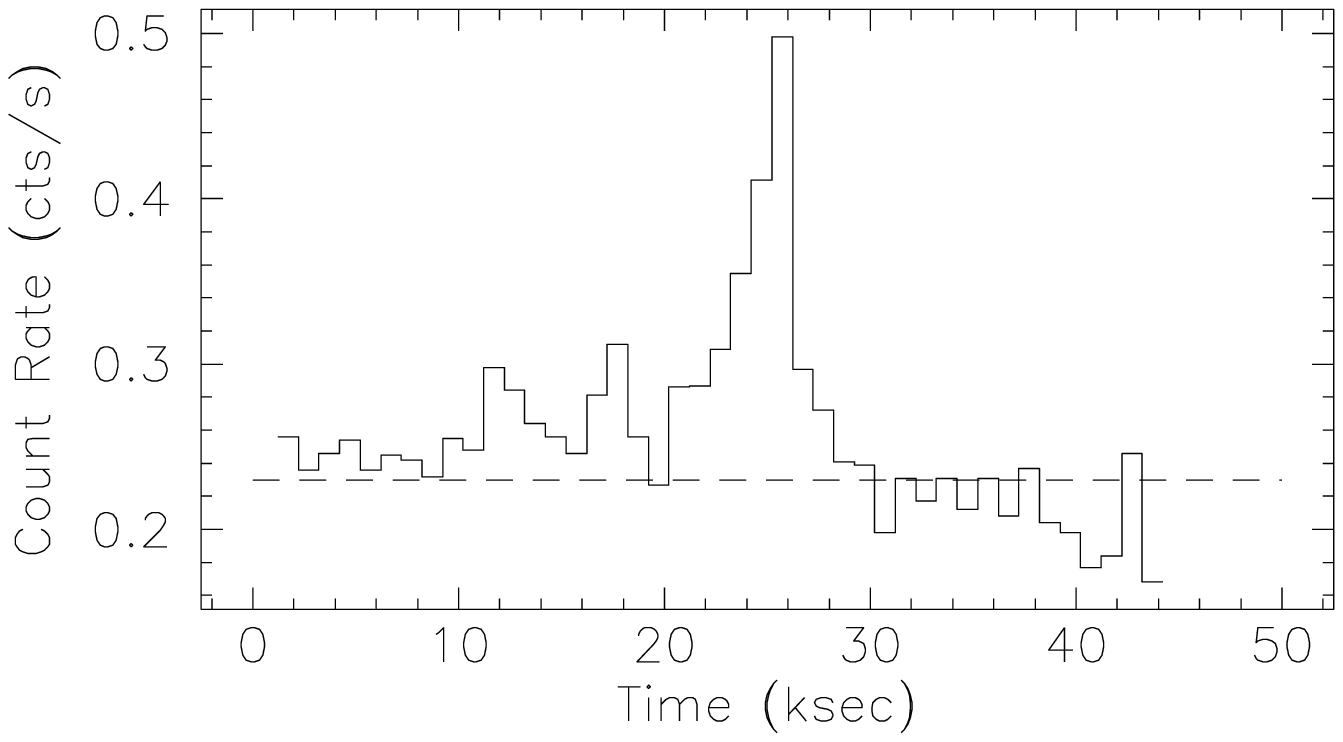}
\includegraphics[totalheight=2.3in]{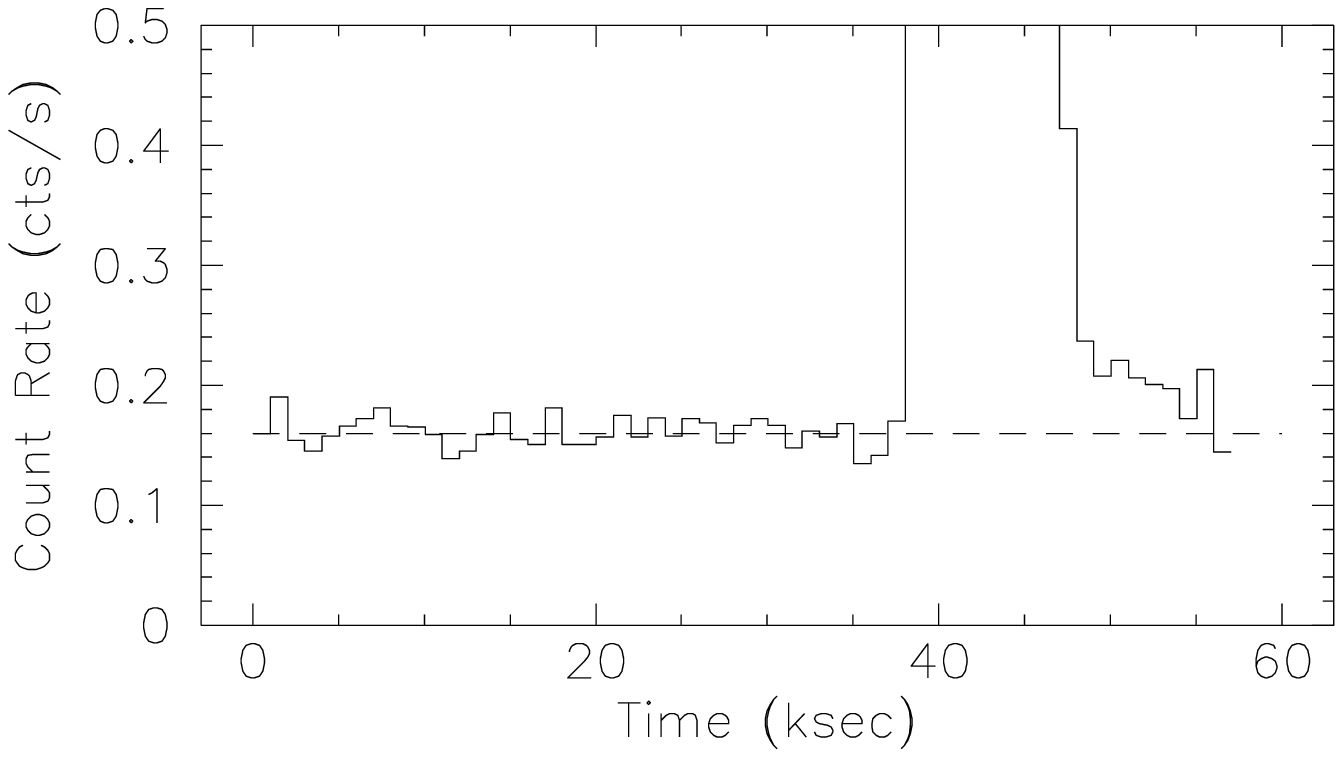}
\caption{(a) Lightcurve from the the first MBM12 observation between
  2.5-7 keV, with a dashed line showing our best-fit quiescent rate of
  0.22 cts/s. (b) Same, from the second observation, with a quiescent
  rate of 0.16 cts/s.  \label{fig:ltcrvwsrcs}}
\end{figure}

We used CIAO 3.1 tools to process the observations, along with CALDB
2.27.  Our result depends crucially on the background measurement,
which must be done indirectly since the source fills the field of
view.  The two most important steps are removing flares and point
sources.  We followed a procedure similar to that described in
\citet{Markevitch03}, although as the data were taken in FAINT mode,
VFAINT filtering was not possible.  For the purposes of source-finding
only, we merged the two observations using the {\tt align\_evt}\
routine\footnote{http://asc.harvard.edu/cont-soft/software/align\_evt.1.6.html}.
We then ran {\tt celldetect}, requiring $S/N > 3$, which found 16
sources (including the well-known source XY Ari).  We excluded all
these sources, using circles of $15''$\ radius (except XY Ari, where
we used a $30''$\ radius circle).  This procedure eliminated $15\%$\
of all events, with XY Ari alone accounting for $11\%$, and removed
$5.0\%$\ of the area leaving a total field of view of 67 arcmin$^2$.
We then made a lightcurve of the ACIS-S3 events between 2.5-7 keV for
each observation, as shown in Figure~\ref{fig:ltcrvwsrcs}.  We
constructed a histogram of the observed rates and fit it with a
Gaussian, representing the quiescent rate, plus a constant to roughly
model the flares.  We obtained good fits in both cases which showed
that the quiescent level in the first observation was 0.22 cts/s while
in the second it was 0.16 cts/s.  The quiescent rate in the first
observation is significantly above those shown in
\citet{Markevitch03}, which ranged from 0.1-0.16 cts/s.  These
fluctuations in the background counting rate are likely due to protons
which are trapped in the Earth's radiation belts or directly from
Solar flares.

Following \citet{Markevitch03}, we removed all times when the count
rate was not within 20\% of the average rate for the second
observation.  The maximum allowed rate is therefore 0.192 cts/s, which
limits the first observation to $2.2$\,ksec of ``good time,'' too
little data to be useful.  This exclusion was not capricious.
Although the first MBM12 observation was $\sim 40\%$ brighter than the
average background rate for ACIS-S3, we expended substantial effort
attempting to model it.  All our attempts required adding what
amounted to arbitrary terms at the same energies as our emission
lines, so no conclusive results were possible.  Therefore, we
reluctantly excluded it from the remaining analysis.  After filtering
the second observation, we were left with 38.2 ksec of usable data.

To cross-check these results, we followed \citet{Markevitch03} and
compared the total counts in selected energy bands to the total counts
seen with PHA values between 2500 and 3000. PHA values in this range
correspond to energies above 10 keV, which are outside \chandra's
range and thus effectively measures the ``particle background'' rate.
Table~\ref{tab:PHAratio}\ shows the results for our MBM12 observations
and one of our background datasets, ObsID 62850 (see
\S\ref{subsec:bkgnd}).  In Table~\ref{tab:PHAratio}\ we show the
results from the filtered Obi1 data as well as the 2.2 ksec of
``good'' data from Obi0 as well as values from all of Obi0.  Ratios of
the number of events in these bands to the number of events with PHA
values between 2500 and 3000 are given as well.  The unfiltered Obi0
data show very high ratios, suggesting contamination from some source
besides the normal particle background.  Conversely, the Obi1
observations of MBM12 show slightly higher ratios compared to the
ObsID 62850 ratios, consistent with a weak source such as the diffuse
X-ray background.  The ratios seen for ObsID 62850 are similar to the
average values observed by \citet{Markevitch03} for the EHM
observations.

\begin{table*}
\caption{Total counts and ratios in selected bands\label{tab:PHAratio}}
\begin{tabular}{lllllllll}
\hline \hline
Band & \multicolumn{2}{c}{\sl ObsID 62850} & \multicolumn{2}{c}{\sl
  MBM12 Obi1} & \multicolumn{2}{c}{\sl ``good'' Obi0} &
\multicolumn{2}{c}{\sl ``all'' Obi0}\\   
               & Counts & Ratio  & Counts & Ratio  & Counts & Ratio  & Counts & Ratio\\ \hline
PHA (2500-3000)& 36197  & 1.000  & 21735  & 1.000  & 1407  & 1.000 & 75925  & 1.000\\
0.5-2.0 keV    &  4762  & 0.132  &  4015  & 0.185  &  314  & 0.223 & 46387  & 0.611\\ 
2.0-7.0 keV    &  8545  & 0.236  &  7492  & 0.344  &  513  & 0.365 & 63693  & 0.839\\
5.0-10.0 keV   & 22742  & 0.628  & 14877  & 0.684  &  978  & 0.695 & 96311  & 1.269\\ \hline
\end{tabular}
\end{table*}

\subsection{Background\label{subsec:bkgnd}}

Since the LB fills the field of view in all normal \chandra
observations, determining the true background is non-trivial.  There
are three types of \chandra observations which contain only
instrumental or near-Earth background: the Dark Moon observation
described in \citet{Markevitch03}, the Event Histogram Mode (EHM) data
taken by ACIS during HRC-I observations and also described in
\citet{Markevitch03}, and the 'stowed ACIS' observations first
described in \citet{Wargelin04}.  These latter observations (to date,
ObsID 4286, 62846, 62848, and 62850) were done with ACIS clocking the
CCDs in ``Timed Exposure'' mode and reporting event data in the VFAINT
telemetry format, and with ACIS in a 'stowed' position at which ACIS
receives negligible flux from the telescope and the on-board
calibration source.  We used data from ObsIDs 62848, 62848, and 62850
as our background datasets.  Together these datasets represent 144
ksec of observations, although they were treated independently in our
fits.  ObsID 4286 is less than 10 ksec and was not used.  Between our
observations and the last of these observations (in December 2003)
there was little change in the overall background rate\footnote{see
http://cxc.harvard.edu/ccw/proceedings/index.html/presentations/markevitch/}.

We fit the MBM12 data from Obi1 simultaneously with stowed ACIS data
using both a foreground and background model and {\sl Sherpa}'s CSTAT
statistic.  The fits were restricted to the range 0.4 - 6 keV, since
above 6 keV the background appears to rise due to the near-constant
cosmic ray background combined with the falling mirror response, while
below 0.4 keV the background rises and falls in a highly variable
fashion.  Our foreground model consisted of two unabsorbed delta
functions for the \ion{O}{7} and \ion{O}{8} lines, plus an absorbed
power law and thermal component to represent the cosmic X-ray
background and the distant hot Galactic component.  We did not include
a thermal component with $T \sim 10^6$\,K to represent the LB
emission, as nearly all of that emission would be below 0.4 keV,
except for the oxygen lines.  The absorption was allowed to vary
between \NH = $3.6-8.4\times10^{21}$\,cm$^2$ for both components; the
best-fit value was $6\times10^{21}$\,cm$^2$\ although our results were
not particularly sensitive to this value.  We used the values from
\citet{Lumb02} for the power-law component ($\Gamma = 1.42\pm0.03$,
normalized to $8.44_{-0.23}^{+2.55}$\ photons
cm$^{-2}$s$^{-1}$keV$^{-1}$\ at 1 keV), as well as for the temperature
of the thermal component ($0.2\pm0.01$\,keV).  The normalization on
the thermal component was allowed to vary, since significant variation
has been seen for this absorbed hot gas \citep{KS00}.  The \ion{O}{8}
Ly$\alpha$\ line position was set at 0.654 keV.  However, since the
\ion{O}{7} emission is from a triplet of lines whose dominant member
is unknown, we allowed the centroid of the line complex to float
within a range of \ion{O}{7} line positions between the forbidden line
at 0.561 keV and the resonance line at 0.574 keV.  We found that the
sharp rise seen below 0.5 keV in both the source and background could
be fit using a low-energy Lorentzian, and also included delta
functions at 1.78 keV (Si-K) and 2.15 keV (Au-M) to fit the
particle-induced fluorescence seen at these energies.  Finally, the
particle-induced continuum background was modeled as a line with slope
and offset, which was not folded through the instrumental response.

\section{Results}

\begin{figure*}
\includegraphics[totalheight=2.3in]{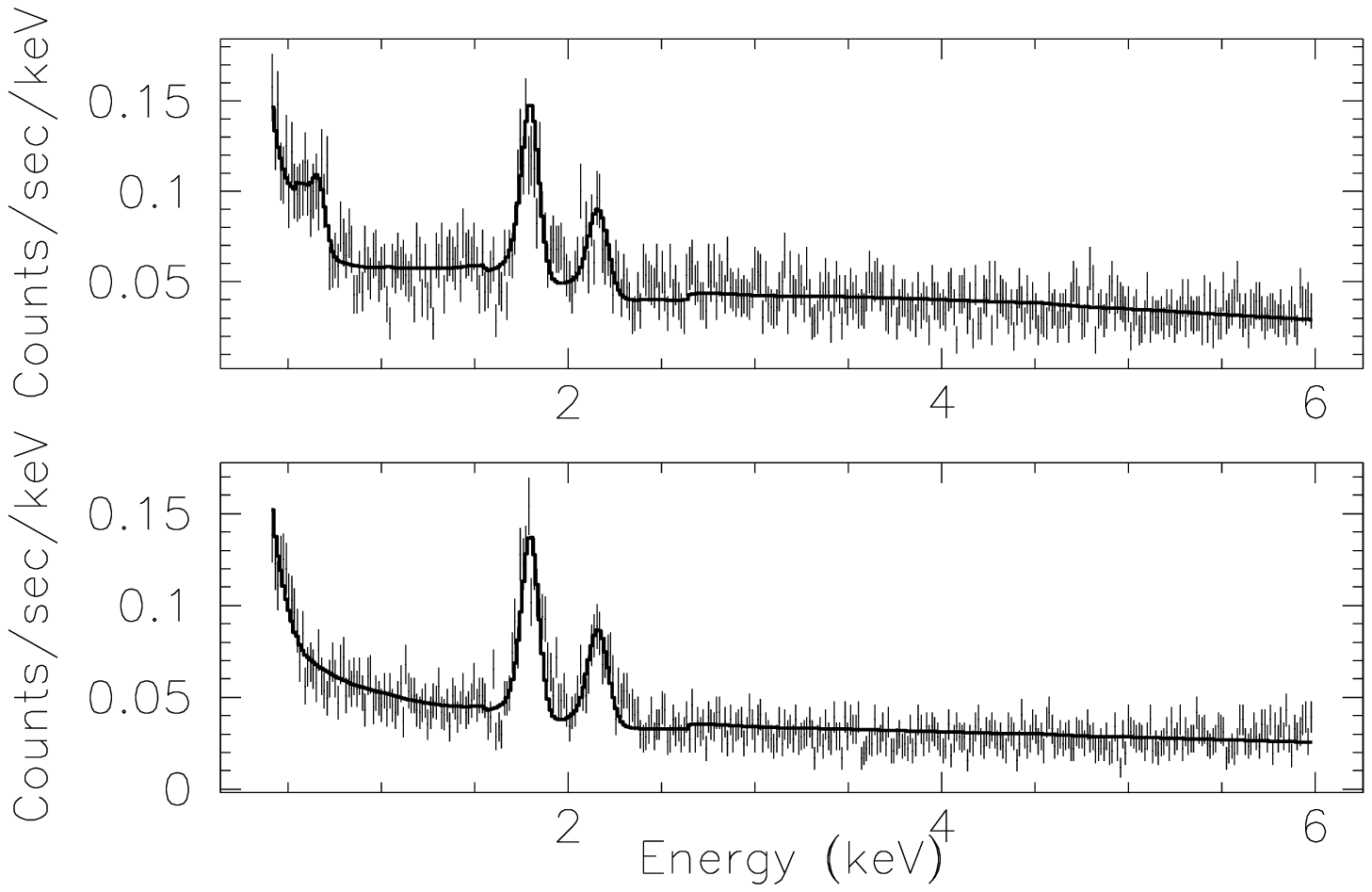}
\includegraphics[totalheight=2.3in]{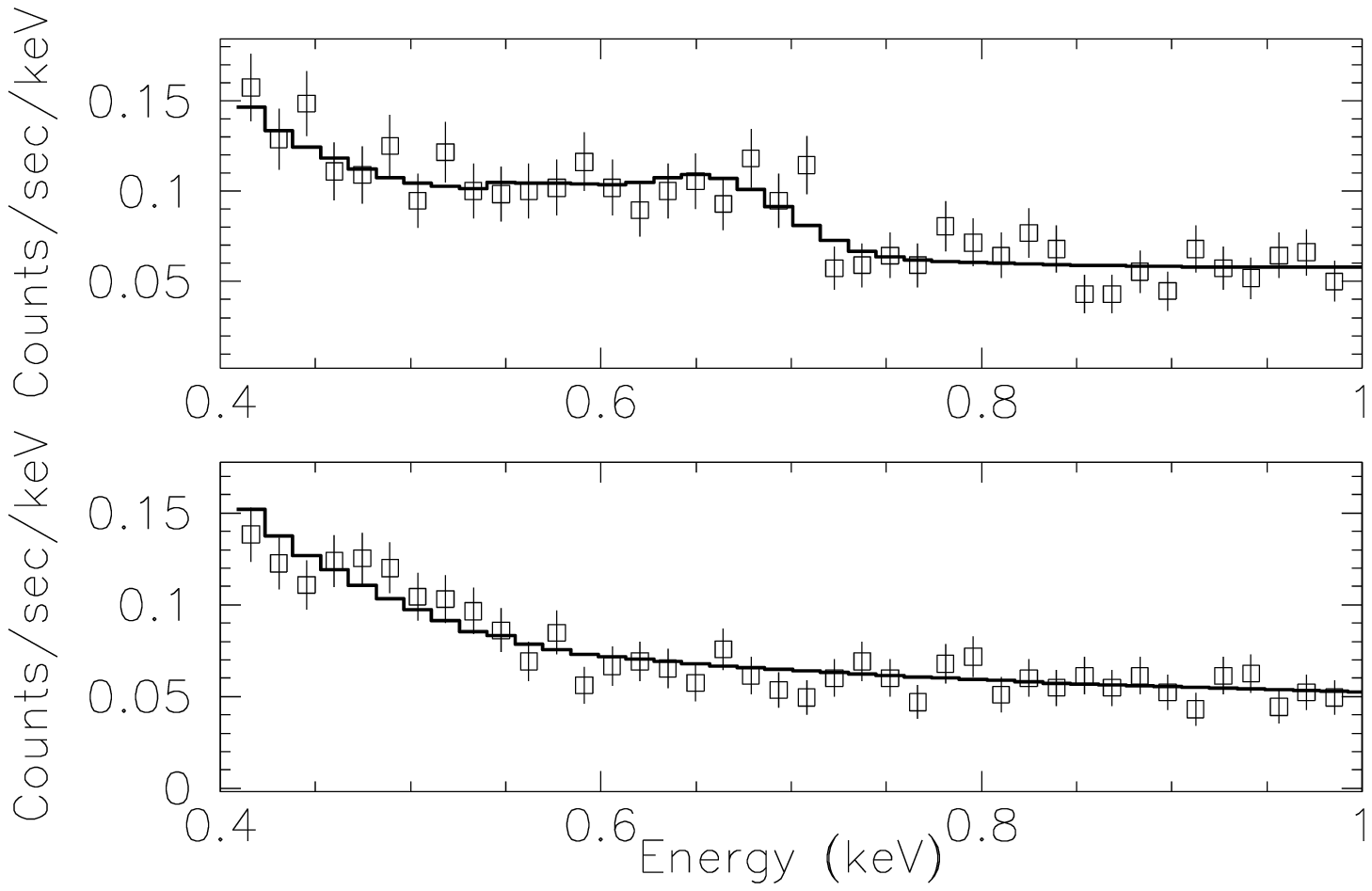}
\caption{[Top Left] Best-fit spectrum for MBM12 observation [Bottom
    Left] Best-fit background spectrum from ObsID 62850; note the
    fluorescence lines at 1.78 keV and 2.15.  [Top Right] Best-fit
    spectrum from MBM12 between 0.4-1 keV, showing the strength of the
    oxygen lines.  [Bottom Right] Same showing the background
    fit.\label{fig:MBM12fit}}
\end{figure*}

Since we used the standard model of the diffuse X-ray background, we
were not surprised to find a good fit to the data.  The source model
had only three significant parameters: the \ion{O}{7} and \ion{O}{8}
line fluxes, and the normalization on the absorbed Galactic thermal
component.  The total absorbed flux from this last component was only
$F_X$(0.4-6 keV) = 0.066 photons cm$^{-2}$s$^{-1}$sr$^{-1}$,
significantly less flux than from either of the two oxygen lines.  Our
results for the line emission are shown in Figure~\ref{fig:MBM12fit}
and Table~\ref{tab:MBM12fit}.  Table~\ref{tab:MBM12fit} also lists
\ion{O}{7} and \ion{O}{8} fluxes measured at high Galactic latitude
with ASCA \citep{Gendreau} and the XQC \citep{XQC02}.  In addition,
line fluxes due to heliospheric charge exchange \citep[labeled
SCK04]{SCK04} and geocoronal charge exchange \citep[labeled ``Dark
Moon'']{Wargelin04} (see \S\ref{subsec:cx}) are listed, along with the
$2\sigma$\ upper limits from the ROSAT observations of MBM12 (SMV93).
The listed errors are $1\sigma$, except for the \citet{Wargelin04}
data where the 90\% likelihood range is shown.  The SMV93 limits
assume all the emission is from that one line, since ROSAT could not
resolve \ion{O}{7} from \ion{O}{8}.  When we allowed the \ion{O}{7}
line to float between the forbidden and resonance line energies, we
found that any value would lead to acceptable fits.  The statistics
and CCD resolution could not distinguish between either the low or
high energy end, so our results are shown for an assumed ``average''
line position of 0.57 keV.  Our best-fit \ion{O}{7} flux is consistent
with the high-latitude \citet{Gendreau} results.  However, the
\ion{O}{8} emission is significantly larger than the results of
\citet{Gendreau} or \citet{XQC02}.  The strength of the \ion{O}{8} is
not strongly dependent on the details of background subtraction, since
it is far from the rapidly rising background found at low energies on
ACIS-S3.  We conclude that this measurement is correct, but
contaminated with Solar system emission from charge exchange.  Other
measurements in Table~\ref{tab:MBM12fit} show that charge exchange can
easily swamp any LB emission.  For example, the \citet{Wargelin04}
``bright'' results from the dark Moon (from their Table 7) are
significantly larger than either our results or those of
\citet{Gendreau}, and the same is true of the SCK04 results.

\begin{table}
\caption{Oxygen line emission (in LU)\label{tab:MBM12fit}}  
\begin{tabular}{lllllll}
\hline \hline
Ion       & This work            & ASCA        & XQC       & SCK04
&Dark Moon&SMV93 \\ \hline
\ion{O}{7}&$1.79\pm0.55$& $2.3\pm0.3$ &$4.8\pm0.8$& $7.39\pm0.79$& 6.5-13.6         &$<7.1$\\
\ion{O}{8}&$2.34\pm0.36$& $0.6\pm0.15$&$1.6\pm0.4$& $6.54\pm0.34$& 2.7-6.1          &$<3.6$\\ \hline
\end{tabular}
\end{table}

\section{Discussion}

Our original goal was to determine the age of the LB by measuring the
ratio of the \ion{O}{7} and \ion{O}{8} emission lines.  In
equilibrium, the \ion{O}{8} emission would be negligible.  A number of
non-equilibrium models, however, predict detectable \ion{O}{8}.
\citet{SC01} considered a range of models involving a series of
supernova explosions.  If the LB is still evolving 3-7 Myr after the
last explosion, their Figure 13b suggests that the \ion{O}{8} would
still be detectable relative to the \ion{O}{7} emission\footnote{Note
that in \citet{SC01}, Figures 11, 12, 19, and 20 are incorrect; they
should be reduced by a factor of $4\pi$, due to an error by the
author.}.  Alternatively, \citet{BS94, B01} would also imply a
detectable amount of \ion{O}{8} emission (see \S\ref{subsec:BS}).
Solar flares, the variable response of the ACIS-I, and complicated
background of the ACIS-S3 CCD have proved significant but not
insurmountable obstacles.  However, the strong \ion{O}{8} line is a
much larger problem, since it is significantly larger than has been
measured at high latitudes which include both the LB and Galactic halo
emission.  As stated above, it seems likely that our results have been
contaminated by charge exchange with the solar wind.

\subsection{Charge Exchange\label{subsec:cx}}

Charge exchange, as first noted by \citet{Cox98}, is a serious
complication for diffuse soft X-ray background studies.  Charge
exchange creates X-rays when electrons jump from neutral material
(usually hydrogen or helium atoms) to excited levels of highly ionized
atoms.  Charge exchange between neutral H and \Oe, for example, can
create \ion{O}{7} emission.  The highly charged ions come from the
solar wind and coronal mass ejections (CMEs).  The neutral material
can come from either the geocorona or the interstellar medium as it
flows into the heliosphere; see \citet{Cravens00} for an overview of
this mechanism.  \citet{CRS01} show that the so-called ``Long Term
Enhancements'' (LTEs) observed during the ROSAT sky survey, which were
apparent brightenings of the diffuse X-ray sky lasting days to weeks,
can be explained by X-ray emission from charge exchange between the
solar wind and either the geocorona or the interstellar medium, or
both.

\citet{Wargelin04} detected a strongly time-dependent \ion{O}{7} line
along with weaker \ion{O}{8} emission from \chandra observations of
the dark Moon.  The source of these X-rays is almost certainly charge
exchange between solar wind ions and the geocorona. The time
variability of the solar wind makes the time-dependence of the X-ray
emission understandable.  Detailed spectral models of charge exchange
emission have been developed \cite[e.g.][]{Kharchenko01}.
\citet{Wargelin04} used these calculations to model the \ion{O}{7} and
\ion{O}{8} geocoronal emission as a function of the solar wind oxygen
ion flux (which can be estimated from ACE measurements).  Their model
estimates the surface brightness in each line as
\begin{equation}
L_S =  {{v_c n_p f_O y_{il} \sigma_i n_{n0}}\over{4\pi}} 5 R_E \big({{10
    R_E}\over{r_{min}}}\big)^2 
\end{equation}
where $v_c$\ is the solar wind velocity, $n_p$\ the proton density,
$f_O$\ the relative abundance of oxygen, $y_{il}$\ the line yield per
charge exchange, $\sigma_i$\ the charge exchange cross section, and
$n_{n0}$ the neutral particle density at 10 Earth radii ($R_E$).
$r_{min}$\ is the geocentric distance to the edge of the magnetosphere
or the spacecraft position, whichever is farther.  Some of these
parameters are measured by the ACE satellite\footnote{Data available
at http://www.srl.caltech.edu/ACE/}.  This model predicts that the
\ion{O}{8}/\ion{O}{7} ratio due solely to charge exchange is $1.36
(n_{O^{+8}}/n_{O^{+7}}) + 0.14$, where $n_I$\ is the density of ion I.
This result uses values from Tables 5 and 6 of \citet{Wargelin04} and
including the \ion{O}{7} K$\beta$\ line in the \ion{O}{8} emissivity,
since they would be blended.  In a typical slow solar wind,
$n_{O^{+8}}/n_{O^{+7}} \approx 0.35$, implying a ratio of 0.616
(SCK04).  Unfortunately, ACE does not yet routinely provide
measurements of $n_{O^{+8}}/n_{O^{+7}}$, as it does with
$n_{O^{+7}}/n_{O^{+6}}$.  During Obi1, $n_{O^{+7}}/n_{O^{+6}}$\ was
$\sim 0.7$, approximately midway between the average value for the
slow solar wind ($\sim 0.3$) and the value of $\sim 1.4$\ measured
during the brightest dark Moon observations.

Going beyond the Earth-Moon system, the time-variable effects of
charge exchange in the broader heliosphere were measured by SCK04.
They used a series of four {\sl XMM-Newton}\ observations of the
Hubble Deep Field North (a high-latitude patch of sky devoid of bright
X-ray sources) to measure \ion{O}{7} and \ion{O}{8} emission and
correlated their results with solar wind observations from ACE.  They
found that three out of four observations (as well as part of the
fourth) agreed with the standard diffuse X-ray background model, and
that during these times the proton flux and
$n_{O^{+7}}$/$n_{O^{+6}}$\ ratios in the solar wind were
$1.5-3.2\times10^8$\,cm$^{-2}$s$^{-1}$\ and $0.15-0.46$, respectively.
However, part of the fourth observation showed substantially brighter
\ion{O}{7} and \ion{O}{8}, as well as sharp increases in both the
proton flux (to $8.0\times10^{8}$\,cm$^{-2}$s$^{-1}$) and the
$n_{O^{+7}}$/$n_{O^{+6}}$\ ratio (to 0.99).  Their measured
heliospheric surface brightnesses are given in
Table~\ref{tab:MBM12fit}.

Our viewing geometry for MBM12 during Obi1 is such that we would
expect an effect due to charge exchange, especially for \ion{O}{7}.
The heliospheric ``downstream'' direction, with respect to the Sun's
motion through the Local Cloud is centered around ecliptic coordinates
$74.5^{\circ}, -6^{\circ}$\ \citep{LB92}, which corresponds to
December 5 for the Earth's orbital position.  MBM12 ($47.6^{\circ},
+3^{\circ}$ in ecliptic coordinates) is not far from this direction,
so when Obi1 was done on August 17 it should have been viewing a
sparse column of neutral heliospheric gas--mostly He, since the H
would be largely ionized as it flowed by the Sun.  From Table 9 of
\citet{Wargelin04}, we therefore expect, for a typical slow solar
wind, a total ROSAT surface brightness of $\sim 470$\,cts
s$^{-1}$sr$^{-1}$.  The geocoronal component should be roughly $\sim
1/4$th of this, as the viewing direction is (roughly) through the
flank of the magnetopause, again from Table 9 of and Figure 7 of
\citet{Wargelin04}.  We therefore expect the total surface
brightnesses will be $\sim 5\times$\ the geocoronal model values in
Table 8 of \citet{Wargelin04}, or 1.4 and 0.56
ph cm$^{-2}$s$^{-1}$sr$^{-1}$\ for \ion{O}{7} and \ion{O}{8},
respectively.  Given the uncertainties, and the expectation that
charge exchange emission does not account for the entire soft X-ray
background observed by ROSAT \citep[but see][]{Lallement04}, the
\ion{O}{7} prediction matches our result well, but the \ion{O}{8} is
still anomalously high for a typical slow solar wind alone.

However, in addition to the solar wind, it should be noted that there
were a number of halo CME events during 2000 July\footnote{see catalog
at
ftp://lasco6.nascom.nasa.gov/pub/lasco/status/LASCO\_CME\_List\_2000)}.
These would have the effect of mixing into the Earth's side of the
interplanetary medium some CME plasma that may well have a
substantially higher ionization state.  These vary from event to
event, and there are only a few events which are suitable for study
both in the near sun environment with SOHO and \textit{in situ}\ with
a solar wind ion charge state spectrometer.

As an example, one such event was observed in 2002 November and
December \citep{Poletto04}. The UVCS data measured remotely by the
SOHO observatory at 1.7 solar radii above the solar limb show lines of
Fe XVIII, and the SWICS data measured \textit{in situ}\ by {\sl Ulysses}\ also
show high ions of iron (Fe$^{+16}$\ being prominent). This leads these
authors to conclude that, for this CME event, the ionization state is
characteristic of a "freezing in" temperature in excess of 6 MK. At
such temperatures, oxygen is fully ionized.

\citet{Cravens00} argues that solar wind charge exchange with the
neutral interstellar medium (ISM) mostly occurs at distances of a few
AU from the sun because of the depletion of neutral gas near the Sun.
Since 100 km\,s$^{-1}$\ translates to about 1.7 AU per month, and CME
fronts range from about 20 to over 2000 km\,s$^{-1}$\,\citep{Webb02},
we are not concerned with a single event but an average over the week
or two prior to the observation.

We therefore conclude that the large \ion{O}{8}/\ion{O}{7} ratio we
observe could be expected from the charge exchange of CME plasma with
the neutral ISM, but that the details are sufficiently complex (and
the data sufficiently sparse) that prediction of this ratio is
difficult to impossible.  We can say, however, that the observed
\ion{O}{7} and \ion{O}{8} provides a (weak) upper limit to the
emission from the LB.

\subsection{Collisional Models} 

What do these observations imply if the observed line emission is due
to electron/ion interactions rather than charge exchange?  In this
case it is relatively easy to calculate the expected contribution from
each atomic process.  \ion{O}{7} emission from any of the triplet of
lines is created by direct excitation (where we include cascades from
excitation to higher levels), electron recombination onto \Oe, or
inner-shell ionization of O$^{+5}$.  \ion{O}{8} lines come from direct
excitation or recombination from bare oxygen ions; K-shell ionization
of \Os does not have a large cross section for emitting \ion{O}{8}.
If the plasma is at or near ionization equilibrium we can ignore
blending from satellite lines or other emission lines in the region of
interest as they should contribute only a small fraction of the total
emission; we consider the far-from-equilibrium case in
\S\ref{subsec:BS}.  For example, at CCD resolution the \ion{O}{7}
$1s3l\rightarrow 1s^2$\ line will blend with the \ion{O}{8}\
Ly$\alpha$\ line.  However, it contributes at most 6\% of the flux of
the \ion{O}{8}\ Ly$\alpha$\ line and so can be ignored to a first
approximation.

\begin{figure}[ht]
\includegraphics[totalheight=2.3in]{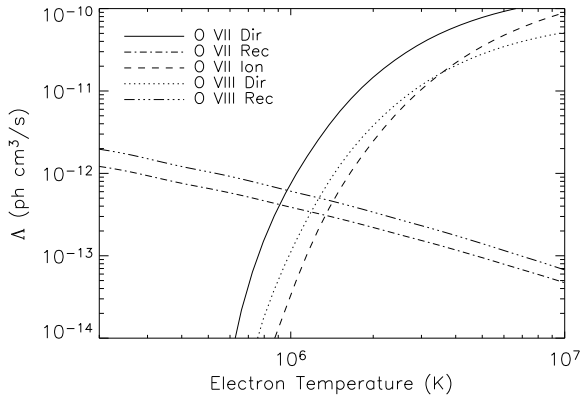}
\caption{$\Lambda$\ curves as a function of electron temperature,
  using APEC and data from ATOMDB 1.3.1. The total emission from each
  process is $n_e n_I \Lambda.$ \label{fig:Lambda}}
\end{figure}

With these assumptions, the observed line surface brightness (in LU)
along a particular line of sight can be written as
\begin{eqnarray}
\nonumber L_{S}(\hbox{\ion{O}{7}}) & = & {{1}\over{4\pi}} \int dR
  {{n_e^2(r)}\over{1.2}}[n_{+5}(r)\Lambda_{\hbox{\small \ion{O}{7}}}^{IS}(T)+\\
 &   &  n_{+6}(r) \Lambda_{\hbox{\small \ion{O}{7}}}^{DE}(T) + 
  n_{+7}(r) \Lambda_{\hbox{\small \ion{O}{7}}}^{RC}(T)] \\ 
\nonumber L_{S}(\hbox{\small \ion{O}{8}})  & = & {{1}\over{4\pi}} \int dR
{{n_e^2(r)}\over{1.2}}[n_{+7}(r) \Lambda_{\hbox{\small \ion{O}{8}}}^{DE}(T) +\\
  &  & n_{+8}(r) \Lambda_{\hbox{\small \ion{O}{8}}}^{RC}(T)] 
\label{eq:lambda}
\end{eqnarray}
where L$_S$\ is in LU, $n_e(r)$\ is the electron density at distance
$r$, $n_{+n}(r)$\ is the ion density of O$^{+n}$\ at $r$.  We also
assume that the hydrogen and helium are fully ionized, so $n_e \approx
1.2 n_H$.  $\Lambda_{I}^{IS}(T)$, $\Lambda_{I}^{DE}(T)$,
$\Lambda_{I}^{RC}(T)$\ are the rate coefficients (in cm$^3$\,s$^{-1}$) for
creating an emission line via inner-shell, direct excitation, or
recombination for ion I, and are plotted in Figure~\ref{fig:Lambda}.

Since four different ions may be involved in creating \ion{O}{7} and
\ion{O}{8} emission lines, we cannot uniquely solve for the density,
temperature, and ionization state of the system.  However, we can
easily show that these line fluxes are not commensurate with an
equilibrium plasma.  The \ion{O}{8} flux provides the strongest limit,
as this line is not expected in an equilibrium model for the LB.
Figure~\ref{fig:Lambda} shows that the direct excitation rate for this
line rises rapidly with temperature, so a higher temperature will
produce more \ion{O}{8}.  Based on the \citet{KS00} survey of LB
results (see their Table 1), we assume that the allowed electron
temperature in the LB is $< 2\times10^6$\,K; in equilibrium at this
temperature $n_{O^{+6}}$/$n_{O^{+7}} > 1$.  Figure~\ref{fig:Lambda}
clearly shows that the direct excitation rate for \ion{O}{7} is more
than triple the rate for \ion{O}{8} excitation, which would predict
\ion{O}{7}/\ion{O}{8} $>3$, in strong contrast to our result.
Therefore, either the \ion{O}{8} line emission is not coming from the
LB, or the plasma is recombining, with significantly more O$^{+7}$\ or
O$^{+8}$\ ions than O$^{+6}$\ ions.  We consider some aspects of this
possibility below.

\subsection{Recombining Plasma Models\label{subsec:BS}}

\citet{BS94} proposed a radical new model for the LB suggesting that
it was not in fact ``hot'' but rather the X-ray emission was created
by recombination of highly-ionized atoms with cool ($10^4 -
3\times10^5$\,K) electrons.  Their model assumes the LB was a
pre-existing cavity, probably formed by a earlier supernova explosions
and possibly associated with the Sco-Cen superbubble.  Then inside
this cavity a few million years ago a massive star inside a dense
cloud exploded.  After the SN shock heated and ionized the entire
cloud, the resulting hot ions expanded into the low-density cavity and
cooled adiabatically.  This model was partially inspired by the
relatively high average density of 0.024 cm$^{-3}$\ measured from a
nearby (130 pc distant) pulsar's dispersion measure\citep{Reynolds90}.
If this density is representative of the LB as a whole, a $\sim
10^6$\,K plasma would create far more soft X-rays than are observed.
\citet{BS94} showed that a low-temperature, high-ionization state
plasma with this density could generate the observed 1/4 keV X-ray
emission.  Later papers \citep{B96, B01} have described the model in
more detail and made predictions of the flux in various bandpasses.
This model also naturally explained the existence of the ``Local
Fluff,'' neutral \ion{H}{1} clouds \citep[e.g.][]{Bochkarev87,
Frisch95} which are hard to understand in the context of ``hot''
bubble models.

Although more complicated than the equilibrium isothermal LB model, a
strength of this model is that it does not require a particular
fine-tuning or a complicated mix of hydrodynamic and plasma models.
As noted by \citet{B01}, only some basic input parameters -- cavity
radius R, electron density $n_e$, electron temperature $T_e$, and
evolution time $t$\ -- are needed and a fairly wide range of these
values matched the available data at the time.  

However, FUSE observations \citep{Shelton03} put a surprisingly low
$2\sigma$\ upper limit on the \ion{O}{6} $\lambda\lambda$1032,1038
emission from the LB of 500 and 530 LU, respectively, and 800 LU for
both lines combined.  \citet{Shelton03} noted that this result alone
strongly constrains any recombining plasma model and when combined
with other discrepancies in the \ion{C}{3} emission and the
N(\ion{O}{6}) column density ``practically eliminate[s] this class of
models.''  However, this argument rests on discrepancies in two
unrelated ions, \ion{O}{6} and \ion{C}{3}, leaving open the
possibility that these issues could be finessed with the
correctly-tuned input parameters.  Recombining plasma models are still
being invoked in recent papers \citep[e.g.][]{FBA04, CHIPS}.  The
\ion{O}{6} upper limits are quite stringent, however, and combining
them with our measurements allows a rigorous test of the model using
all relevant oxygen ions as they recombine from fully stripped
O$^{+8}$\ to O$^{+5}$\ and beyond.  Any acceptable recombining plasma
model would have to match the upper limits found for each ion's
emission lines, while also emitting at least some of the observed 1/4
keV band X-rays.

Interestingly, in the case of a purely recombining plasma, the
ionization state of the plasma at any point in its evolution can be
calculated analytically.  If ionization can be ignored, the population
of any ion state can be written as:
\begin{equation}
{{dp^{i}}\over{dt}} = - R_{i} n_e p^{i} + R_{i+1} n_e p^{i+1}
\label{eq:recpop}
\end{equation}
where $p^i$\ is the population of the ion with $i$\ electrons and
$R_{i}$\ is the recombination rate (in cm$^3$\,s$^{-1}$) from the
$i$th ion state to the $i+1$\ state.  In the case of the
fully-stripped ion ($p^0$), the second term in Eq.~(\ref{eq:recpop})
is zero as there is no higher ion.  Equation (\ref{eq:recpop}) then
leads to a linked series of first order differential equations that
can be easily solved.  Assuming that initially the population is fully
stripped ($p^0(0) \equiv 1$), the population of any ion can be written
as:
\begin{equation}
p^{i} = \sum_{j=0}^{i-1} \Big[ {{R_{i-1} f_j^{i-1}}\over{R_i - R_j}}
  (\exp(-R_j n_e t) - \exp(-R_i n_e t)) \Big]
\end{equation}
where $f_j^i$\ equals 1 when $i=j=0$\ and is otherwise defined by the
recursion relation:
\begin{equation}
f_j^i=\cases{ {{R_{i-1} f_j^{i-1}}\over{R_i - R_j}}&if $j < i$ \cr
  -\sum_{j=0}^{i-1} {{R_{i-1} f_j^{i-1}}\over{R_i-R_j}}& if $j=i$\cr }
\end{equation}

To calculate $R_i$\ we used radiative recombination rates from
\citet{VF96} and dielectronic rates from \citet{Romanik88,
RomanikPhD}.  Combining this with Figure~\ref{fig:Lambda}\ we can
calculate the predicted \ion{O}{8} Ly$\alpha$\ and \ion{O}{7} triplet
flux due to recombination for any temperature and density.  We
calculated the \ion{O}{6} flux from electron collisions (recombination
from O$^{+6}$\ was negligible) using the collision strengths from
\citet{GBP00}.

\begin{figure*}[ht]
\includegraphics[totalheight=2.5in]{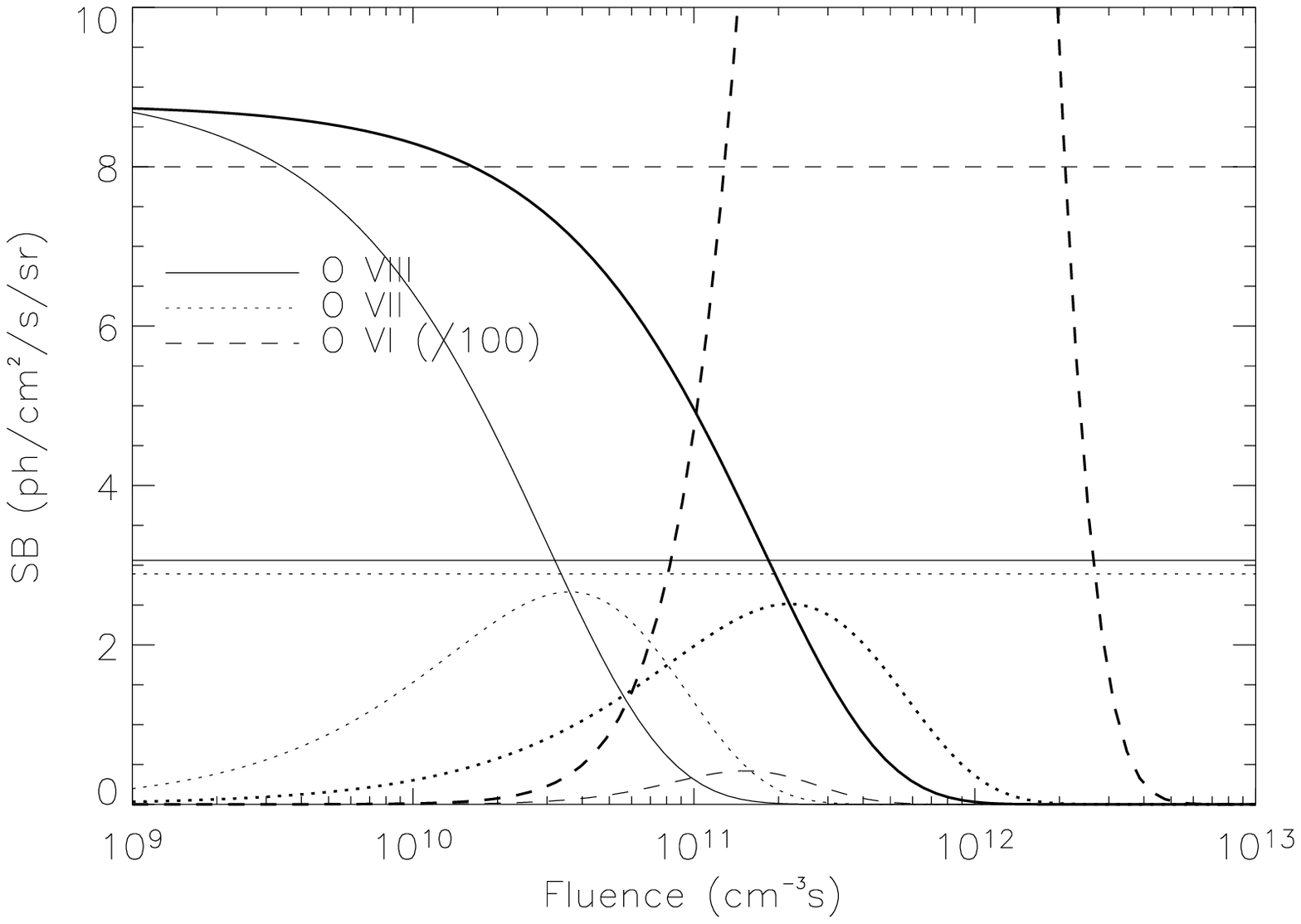}
\includegraphics[totalheight=2.5in]{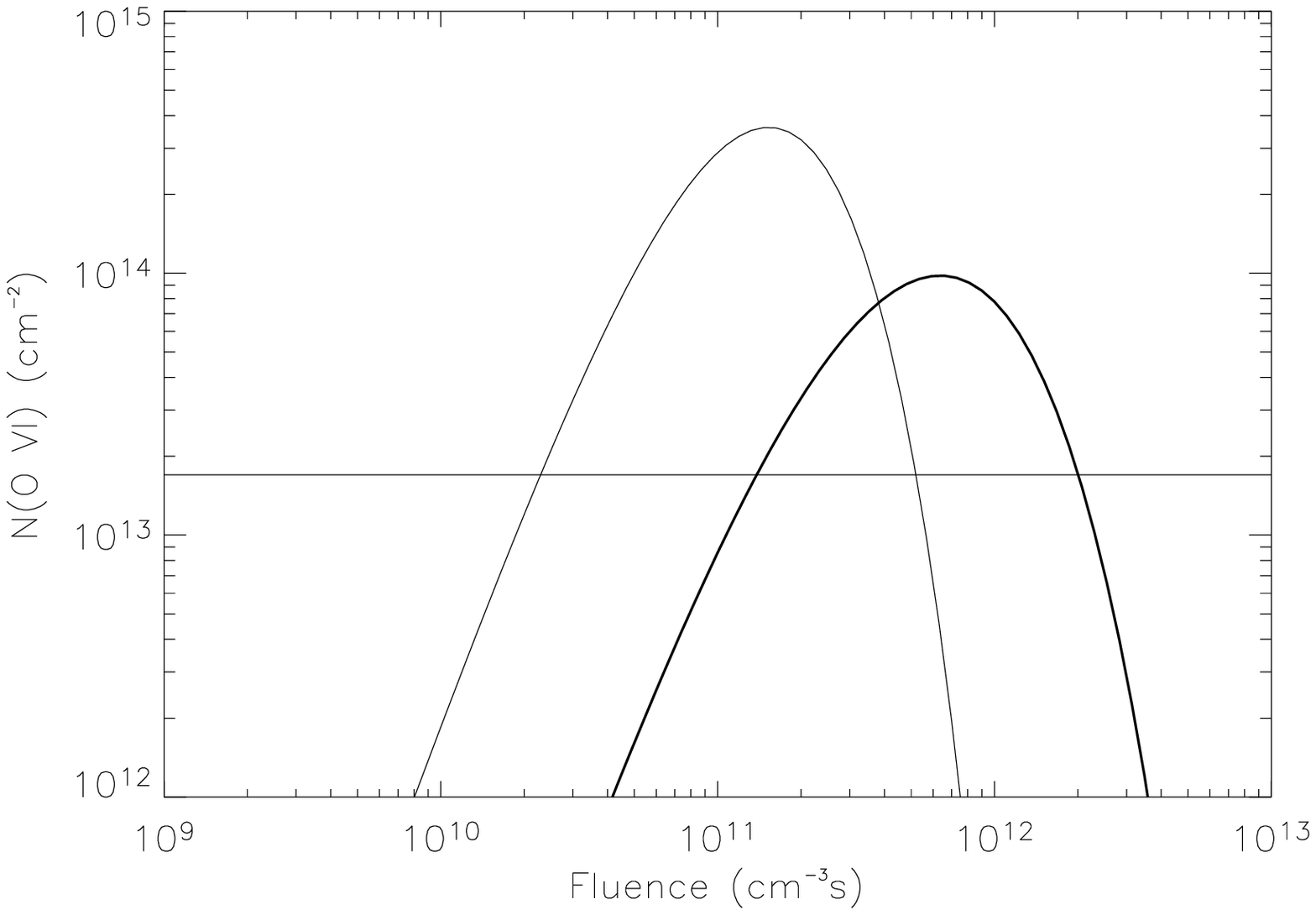}
\includegraphics[totalheight=2.5in]{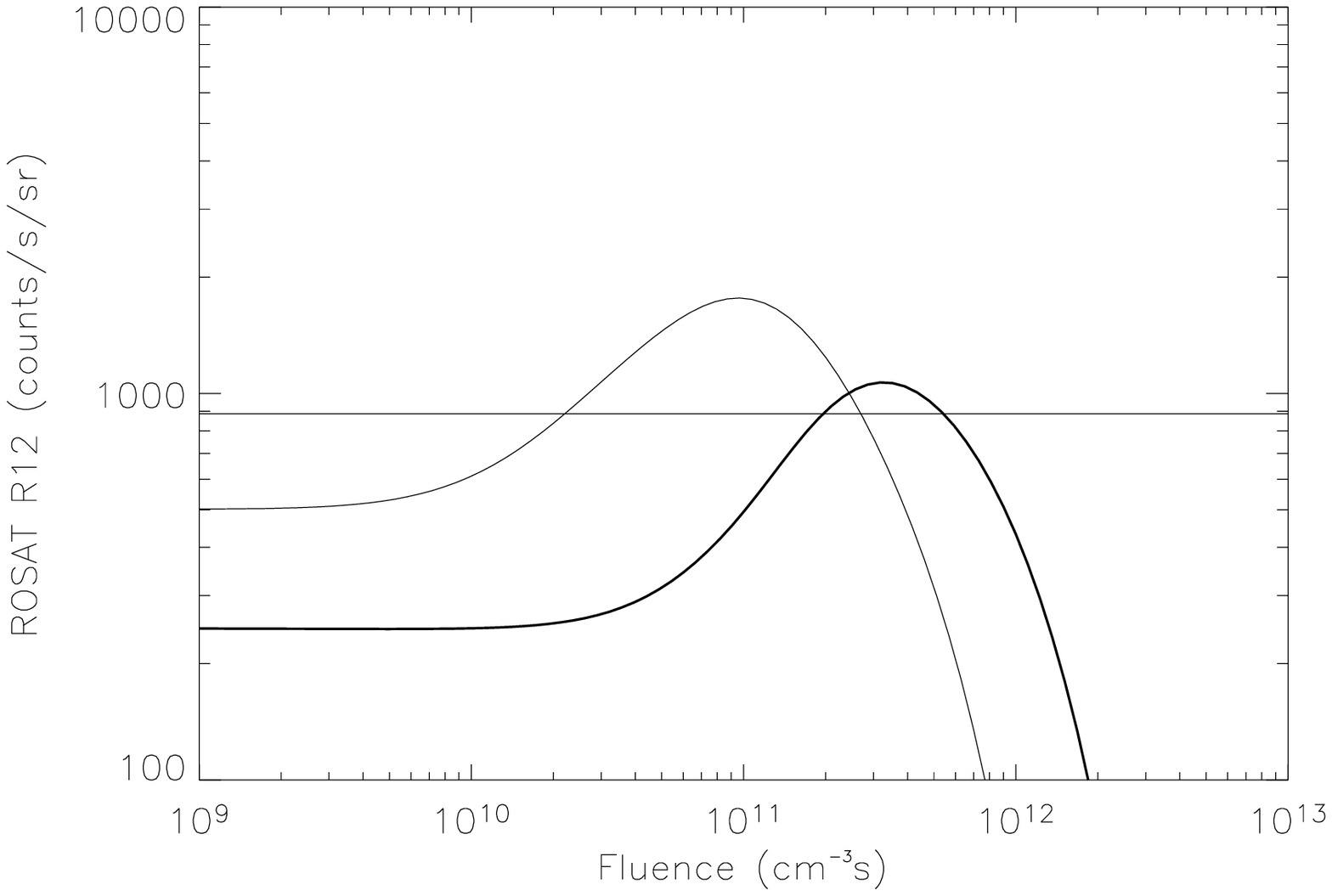}
\includegraphics[totalheight=2.5in]{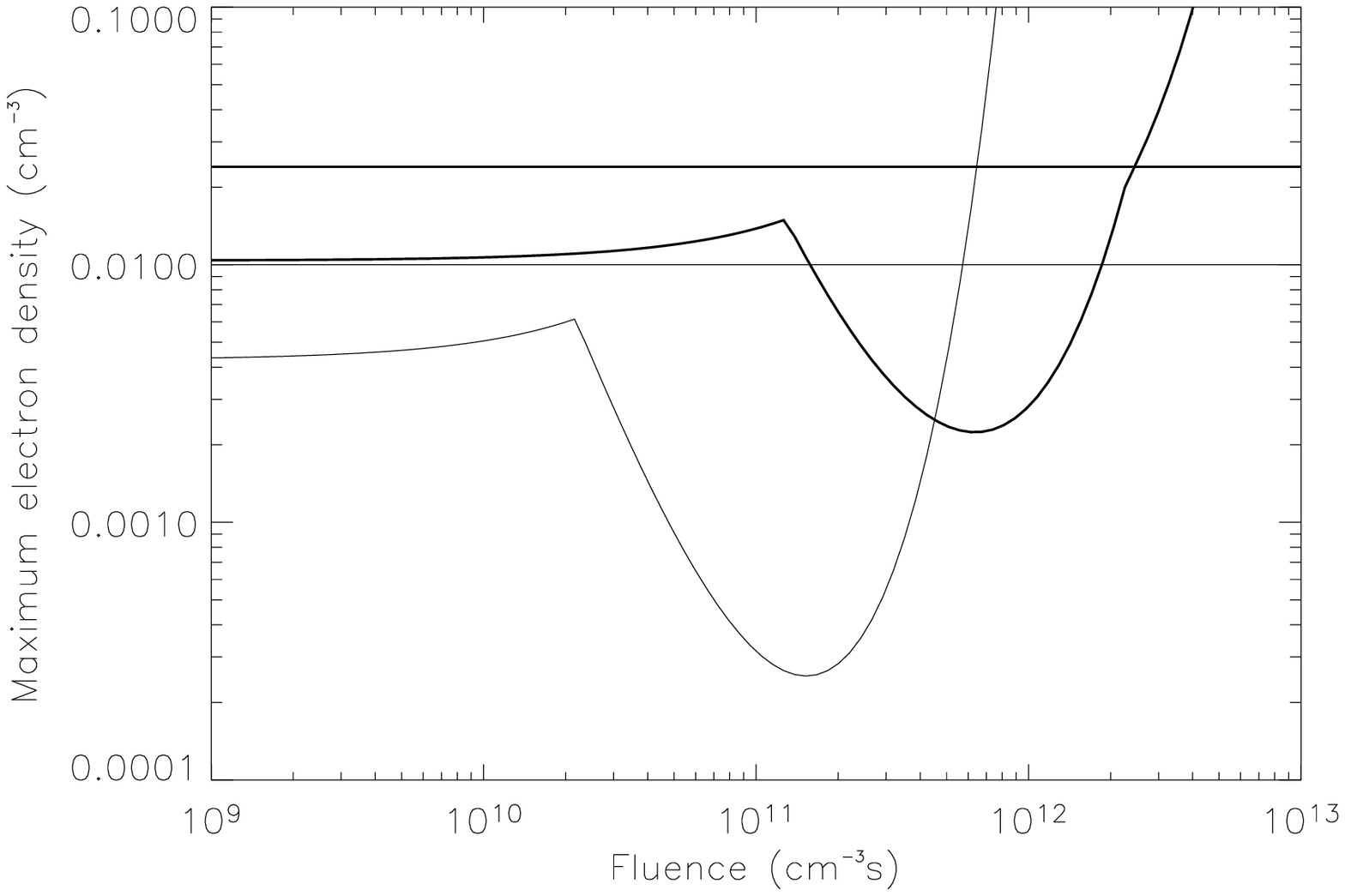}
\caption{[Top Left] The \ion{O}{8} Ly$\alpha$, \ion{O}{7} triplet, and
  \ion{O}{6} doublet emission and observed upper limits for two
  different models as a function of fluence.  Thick curves show a
  model with $R=114$\,pc, $n_e = 0.024$\,cm$^{-3}$, and $T_e =
  3\times10^5$\,K, thin lines are for a model with $R=114$\,pc, $n_e =
  0.01$\,cm$^{-3}$, and $T_e = 2\times10^4$\,K [Top Right] The
  \ion{O}{6} column density in the LB for the same models, along with
  the observed upper limit from the LB. [Bottom Left] The ROSAT R12
  emission from both models, with a line showing our estimated lower
  limit. [Bottom Right] The maximum allowed electron density for a 114
  pc bubble in either model, based only on the \ion{O}{8}, \ion{O}{7},
  and \ion{O}{6} limits.  The effects of each Oxygen ion's limit are
  visible.  Except at large fluences (when the R12 emission is
  negligible), the maximum allowed density is less than the assumed
  model density, showing why these models are excluded.
  \label{fig:recspec}}  
\end{figure*}

However, the upper limits on the \ion{O}{8}, \ion{O}{7}, and
\ion{O}{6} emission and the N(\ion{O}{6}) column density were not
sufficient, as a sufficiently low density would always be allowed (see
Figure~\ref{fig:recspec}[Lower Right]).  The recombining ion model,
however, was initially developed with a relatively high density in
mind and tested against the observed 1/4 keV band emission.  We have
developed an approximate but robust model of the X-ray spectrum from a
recombining plasma to compare to the ROSAT R12 band emission.  This
model considered only emission from dielectronic and radiative
recombination, omitting direct excitation.  In a plasma with low
electron temperature, the radiative recombination continuum appears as
a nearly line-like (width $\sim kT$) feature at the binding energy of
the recombined level.  We therefore assumed that each radiative
recombination led to a photon at the binding energy of the ion; this
ignores cascades which would modify somewhat the exact distribution of
the photons but is reasonably accurate for our purposes.  The
dielectronic satellite lines were calculated using data from
\citet{RomanikPhD, Romanik88}.  We then folded the resulting spectrum
between 50-1000 eV through the ROSAT R12 band response to calculate
the expected emission in this band.

For the \ion{O}{7} and \ion{O}{8} limits we used the $2\sigma$\ upper
limits from the MBM12 observation, despite our strong suspicion (based
on the ACE data and the odd ratio of the two lines) that these are
already contaminated by charge exchange.  In addition, in the case of
\ion{O}{8}\ our $2\sigma$\ upper limit is larger than the value
observed by the XQC \citep{XQC02} over a 1 sr field of view, which
necessarily includes both LB and more distant emission.  Therefore, we
feel this is a very conservative upper limit to the local contribution
from \ion{O}{7} and \ion{O}{8}

Defining an ``upper limit'' to the column density of \ion{O}{6} in the
LB is difficult, and so we attempted to determine a reasonable value
by indirect methods.  \citet{SC94} found through a statistical
analysis that the most likely value for N(\ion{O}{6}) in the LB was
$1.6\times10^{13}$\,cm$^{-2}$.  The {\sl Copernicus}\ data this was
based on \citet{Jenkins78} also shows that no O or B star within 250
pc has a column density greater than this value.  In addition, FUSE
results towards 25 nearby white dwarfs \citep{Oegerle04} show at most
$1.7\times10^{13}$\,cm$^{-2}$, with an average value of $\sim
7\times10^{12}$\,cm$^{-2}$.  We therefore chose N(\ion{O}{6}) $<
1.7\times10^{13}$\,cm$^{-2}$\ as our upper limit to the column density
of \ion{O}{6} through the LB.

We wanted to choose a very conservative value for the minimum R12 band
emission from the LB required by the model.  \citet{BS94} assumed some
of the observed soft X-ray emission came from distant superbubbles,
since ROSAT had recently seen clouds in absorption.  Nonetheless, they
also required some local emission, although the required amount was
somewhat uncertain.  Since then, \citet{KSV97} studied the foreground
X-ray emission by analyzing the shadows seen by ROSAT towards nearby
molecular clouds.  They measured the foreground R12 emission seen
towards 9 clouds, finding a range of 3790 - 6190 counts
s$^{-1}$\,sr$^{-1}$.  This was followed by the exhaustive study of
X-ray shadows by \citet{Snowden00}, who examined 378 clouds seen with
 ROSAT.  Their results showed substantial variation from position to
position, with a mean and standard deviation of 5810$\pm$1960 counts
s$^{-1}$\,sr$^{-1}$.  Based on these results, we decided to require at
least 910 counts s$^{-1}$\,sr$^{-1}$\ in the R12 band, far below the
value found by \citet{KSV97} and $2.5\sigma$\ below the
\citet{Snowden00} mean value.

In Figure~\ref{fig:recspec} we show the predicted \ion{O}{8},
\ion{O}{7}, and \ion{O}{6} emission, the \ion{O}{6} column density,
the R12 emission for two recombining plasma models, and the maximum
allowed electron density based solely on the oxygen ion upper limits
for both models.  The oxygen, neon, and argon abundances were taken
from \citet{Asplund04}, with all other abundances from \citet{AG89}.
The plots are shown as functions of the fluence ($\equiv n_e t$), a
convenient variable as all the rates are proportional to the electron
density.  The thick curves show the results for the model described in
\citet{B01}, with a cavity radius of 114 pc, an electron density of
$0.024$\,cm$^{-3}$, and an electron temperature of $3\times10^5$\,K.
Note how the R12 emission matches the observations over a wide range
of fluences, suggesting that our 1/4 keV emission model agrees with
that used in \cite{B01}.  Whenever the R12 emission reaches and
exceeds the lower limit in this model, though, the \ion{O}{6} emission
exceeds its upper limit. The thin curves show a second model, with the
same radius, an electron density of 0.01 cm$^{-3}$, and an electron
temperature of $2\times10^4$\,K.  In this case there is a small region
with fluence $\sim 10^{10}$\,cm$^{-3}$s where the R12 and \ion{O}{6}
emission pass their lower and upper limits, respectively.  However, in
this range the predicted \ion{O}{7} and \ion{O}{8} substantially
exceed the upper limits towards MBM12.  At later fluences the
predicted \ion{O}{7} and \ion{O}{8} emission drops but the
N(\ion{O}{6}) column density exceeds the observed value.  In sum,
there is no point which simultaneously meets all the upper and lower
limits.

The R12 emission calculation is necessary, since it provides the only
lower limit, but it is also quite complex to calculate and it is
useful to consider what can be derived purely from the simpler oxygen
ions.  Figure~\ref{fig:recspec}[Bottom Right] plots the limit on the
electron density available purely from the oxygen emission and
absorption upper limits.  For example, the \ion{O}{8} emission is
given by
\begin{equation}
{\hbox{\ion{O}{8}}} = n_e n_{+8} {R\over{4\pi}} \Lambda_{\hbox{\small
\ion{O}{8}}}^{RC}(T)
\end{equation}
which can easily be converted into an upper limit on the electron
density since the remaining terms are determined by recombining ion
model.  As can be seen in the model, the limits from the Oxygen ions 
alone require the electron density to be below the assumed density
until very late times or fluences.  

We have examined a wide range of input parameters for the recombining
plasma model.  We considered electron densities up to 0.03 cm$^{-3}$,
electron temperatures in the range $10^4-5\times10^5$\,K and all
cavity radii between 30-300 pc and followed each set of values for
$10^7$\,years, the maximum plausible lifetime of the LB.  Although the
results vary substantially with the input parameters, as shown in
Figure~\ref{fig:recspec}, none of our input parameters lead to
predicted values that can simultaneously match these observational
limits.  In essence, the R12 lower limit requires a minimum density of
highly ionized ions while the tight observational limits on
\ion{O}{6}, \ion{O}{7}, and \ion{O}{8} simultaneously limits the
density of these ions.  We can with confidence dispose of at least the
static recombining plasma model.

\section{Conclusions}

We have observed the nearby molecular cloud MBM12 with \chandra, and
measured the foreground \ion{O}{7} and \ion{O}{8} surface brightness
towards the cloud, which should absorb any distant emission.  Our
observed values are higher than expected, and it appears likely, based
on the ACE satellite data, that charge exchange in the heliosphere or
geocorona has contaminated our results.  We measure a unexpectedly low
ratio of \ion{O}{7}/\ion{O}{8} $= 0.76\pm0.26$, which is hard to
understand in the context of other measurements or most LB models.  In
addition, circumstantial evidence also suggests that our results could
be contaminated by charge exchange.

Despite this limitation, we are able to combine our upper limits with
results from FUSE on \ion{O}{6} emission and absorption and the 1/4
keV ROSAT R12 emission to reject the constant temperature recombining
plasma model originally proposed by \citet{BS94} over a wide range of
input parameters.  Since the sole moderate-resolution spectrum of the
LB taken by DXS showed that the constant temperature equilibrium model
could also be rejected \citep{DXS}, this result implies that more
complex models of the soft X-ray emission from the LB are required.

A number of more complex models have been proposed, however
\citep[e.g.][]{SC01,Avillez03}, and selecting amongst the various
possibilities will require high resolution observations.  If for
example we could resolve the \ion{O}{7} triplet we would be able to
determine if the plasma is ionizing, recombining, or both in different
regions.  Although Astro-E2 will have the necessary resolution, its
effective area-solid angle product is only $\sim
330$\,cm$^2$\,arcmin$^2$, compared to $\sim
20,000$\,cm$^2$\,arcmin$^2$\ for ACIS-S3.  Constellation-X, however,
will be a powerful instrument for understanding the recent history of
our local environment.

We are grateful to Don Cox, Mike Juda, John Raymond, Robin Shelton,
Steve Snowden, and Shanil Virani for helpful discussions.  This work
was supported by the \chandra X-ray Science Center (NASA contract
NAS8-39073) and NASA \chandra observation grant GO0-1097X.

\end{document}